# Controlled Manipulation of Intermediate State in a Type-I Superconductor

Xin-Sheng Gao[1], Qun Wang[1], Ya-Xun He[1], Xing-Jian Liu[1], Jun-Han
Zhang[1], Kang-Hong Yin[1], Jia-Ying Zhang[1], and Jun-Yi Ge[1,2,3*]

[1]*Materials Genome Institute, Shanghai University, 200444 Shanghai, China*
[2]*Department of Physics, Shanghai Key Laboratory for High Temperature*
*Superconductors, Shanghai University, 200444 Shanghai, China*
[3]*Institute for Quantum Science and Technology, Shanghai University, 200444 Shanghai, China*

The intermediate state of type-I superconductors presents a classic paradigm of modulated pattern formation, arising from the competition between short-range attractive and long-range repulsive vortex-vortex interactions. However, direct visualization and, more importantly, active control over the topology and dynamics of these flux structures have remained significant challenges, limiting our ability to manipulate them for fundamental studies and potential applications. Here, using low-temperature magnetic force microscopy, we achieve direct imaging and controllable manipulation of the flux structures in a high-purity tantalum single crystal. We systematically track the evolution of flux morphology—from tubes to stripes—during flux penetration and expulsion, revealing a pronounced topological hysteresis originating from the geometric barrier. Furthermore, we demonstrate precise local control by using the magnetic tip to drag and merge individual flux tubes and to reconfigure entire stripe domains. Under global alternating current (AC) excitation, we discover a reversible stripe-grid-stripe transition, a dynamic reorganization driven by current-induced flux penetration and pinning effects. The corresponding phase diagram shows that the threshold current decreases with magnetic field but increases with AC frequency. Our work establishes a pathway for active flux manipulation in type-I superconductors, revealing rich dynamics and paving the way for flux-based superconducting devices.

## I. INTRODUCTION

In type-I superconductors, the intermediate state (IS) arises from the competition between short-range attractive interactions, originating from the positive surface energy of superconducting–normal interfaces, and long-range repulsive interactions associated with magnetic energy. This competition often leads to the self-organization of complex flux patterns, including tubular and stripe-like structures. To date, the IS has been directly visualized using techniques such as Bitter decoration [1–4], magneto-optical imaging [5–7], and scanning Hall probe microscopy (SHPM) [8]. Extensive studies have shown that the formation and evolution of the IS strongly depend on the magnetic field history. For instance, during the superconducting–normal phase transition, superconducting domains exhibit distinct topological differences near the critical field $H_c$ [9]. Moreover, in the flux penetration process, the presence of a geometric barrier gives rise to the formation of a diamagnetic band [10, 11], which results in a pronounced hysteresis in the transition from flux tube to stripe structures [12].

Previous studies have mainly focused on the transformations of IS under direct current (DC) driving [5, 13, 14]. For instance, Hueben demonstrated that DC currents applied in zero magnetic field can induce normal-state channels with opposite polarities on both sides of an In film [15]. In contrast, studies on ac-driven dynamics remain relatively limited. Menghini et al. demonstrated that applying a periodically modulated magnetic field can drive a transition from superconducting stripes to bubbles [16], further revealing the crucial influence of pinning centers on the morphological evolution of the IS. These results highlight the rich dynamical behavior and complex energy competition underlying the IS of type-I superconductors. Therefore, studying the evolution of the IS under ac excitation is of particular importance for understanding the interplay among flux structures, pinning centers, and geometric barriers.

On the other hand, the continuous advancement of microscopic imaging techniques has made it a reality to locally manipulate flux motion. In type-II superconductors, various mature flux-manipulation techniques have been developed, enabling precise motion, generation, and rearrangement of single vortex or vortex arrays through focused laser beams [17], local heating [18], magnetic forces [19], or mechanical stress [20]. In contrast, studies on flux manipulation in the IS of type-I superconductors remain scarce. Achieving local control of flux structures in type-I superconductors would provide new insights into flux dynamics and pinning mechanisms. Moreover, controllable flux motion also holds great potential for applications in superconducting electronics and flux-based logic systems [21, 22].

Although the IS under DC driving has been extensively studied, the dynamic response of flux structures in type-I superconductors under AC excitation still lacks systematic exploration. Specifically, it remains unclear how the geometric barrier and bulk pinning synergistically govern the underlying topological transition mechanisms. Furthermore, adapting the mature flux-manipulation concepts from type-II superconductors to achieve active control over IS flux structures remains a formidable challenge. In this work, we bridge these research gaps by leveraging the unique dual capabilities of low-temperature magnetic force microscopy (MFM). Unlike the imaging techniques previously employed, MFM not only provides direct, high-spatial-resolution visualization of IS flux structures but also enables active and precise manipulation of normal domain. This study is centered on achieving controllable regulation of intermediate-state structures through multiple approaches. We first track the structural evolution under different magnetic field histories via MFM imaging, and subsequently demonstrate the controlled local manipulation of distinct flux configurations using the MFM tip. Finally, by systematically varying the AC current amplitude, frequency, and external magnetic field, we provide an in-depth investigation into the combined regulatory roles of the Lorentz force, pinning centers, and geometric barriers on the evolution of flux structures.

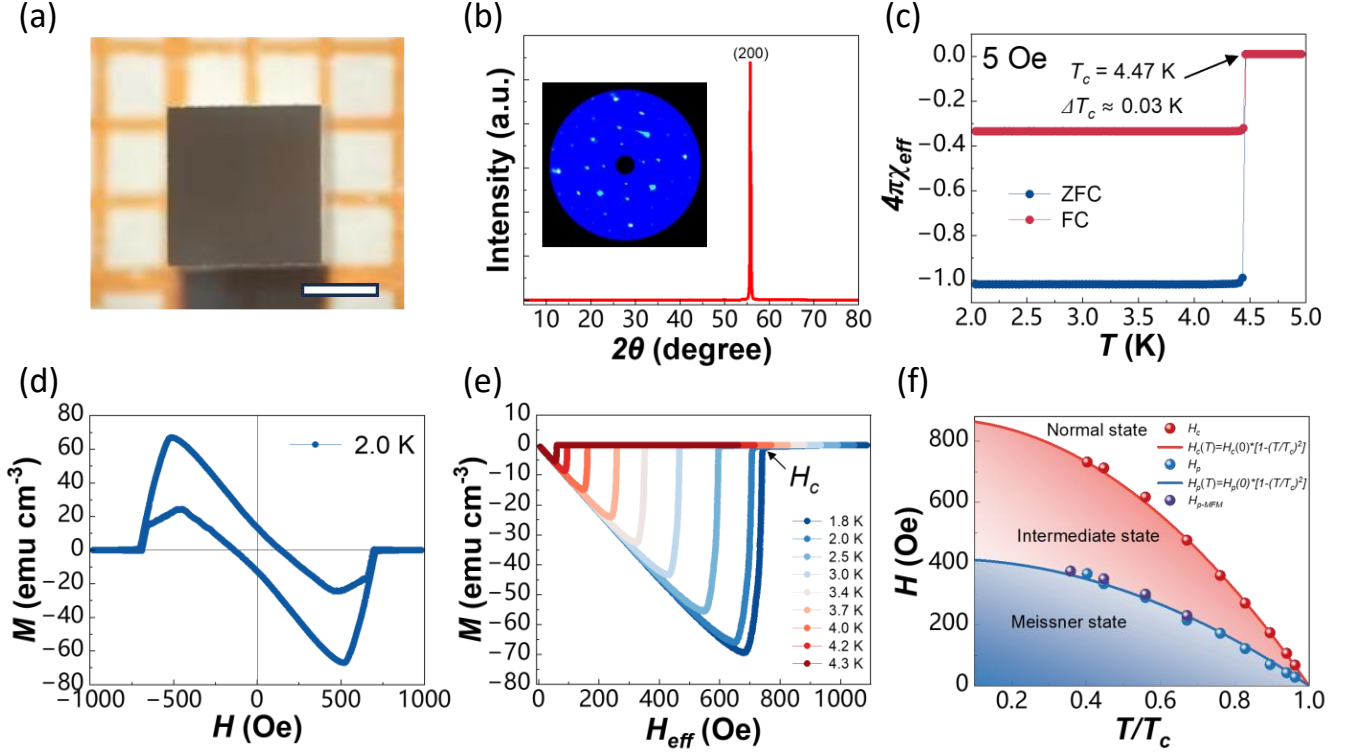

Figure 1. (a) Photograph of the sample. Scale bar: 1 mm. (b) XRD pattern of the Ta single crystal. The upper inset shows the Laue back diffraction pattern. (c) Temperature dependence of the DC magnetic susceptibility in a field of 5 Oe. (d) Magnetic hysteresis loop measured at 2 K. (e) Initial magnetization curves at various temperatures. (f) $H$-$T$ flux phase diagram. The red and blue dots represent $H_p$ and $H_c$ respectively, both taken from the $M$-$H$ curves. The purple dots indicate the $H_p$ extracted from the MFM images, where flux appears in the scanned area. The red and blue lines represent the fitting results for $H_p$ and $H_c$ using the empirical formula $H(T) = H(0)[1 - (T/T_c)^2]$.

## II. MATERIALS AND METHODS

The high-purity Ta single crystal was provided by GermanTech Co., Ltd. Following mechanical polishing, the measured root-mean-square (RMS) surface roughness of the crystal is 0.5 nm. The crystallinity of the sample was verified using X-ray diffraction (XRD) and Laue back-reflection techniques (TDF-3000). Magnetization measurements were carried out with a Quantum Design Magnetic Property Measurement System (MPMS-3). The Ta crystal used for magnetic measurements had dimensions of $2 \times 2 \times 1$ $mm^3$ [Fig. 1(a)]. Flux imaging was performed at 1.6 K using a low-temperature magnetic force microscope (attoDry) in the field-cooled (FC) process. The scanned area is $15 \times 15$ $\mu m^2$, close to the center of the sample. Magnetic contrast was obtained in the non-contact lift mode by recording the phase signal. Both the magnetic and MFM measurements were conducted with the applied magnetic field perpendicular to the (100) plane.

## III. RESULT AND DISCUSSIONS

Figure 1(b) shows the XRD pattern of the Ta single crystal, where a sharp diffraction peak indexed by (200) is observed. The crystal quality was further verified by Laue back-reflection measurements (inset of Fig. 1(b)), which display bright diffraction spots, confirming the high crystalline quality. Figure 1(c) shows the tem-perature dependence of the magnetic susceptibility of the Ta single crystal under an applied field of 5 Oe. A super-conducting transition is observed at $T_c = 4.47$ K with a narrow width of $\Delta T_c \approx 0.03$ K. The demagnetization factor N was calculated to be 0.57 using the formula $N^{-1} = 1 + [3c(1 + ab)]/4a$ [23], and the magnetization data were further corrected for demagnetization effects using $4\pi\chi_{\text{eff}} = 4\pi\chi/(1 - N\chi)$ [24]. The zero-FC (ZFC) susceptibility shows perfect diamagnetism with $4\pi\chi_{\text{eff}} = -1$. Figure 1(d) shows the magnetization loop at 2 K. The narrow magnetization loop indicates the relatively weak pinning of the sample and the related low criti-cal current density. Figure 1(e) shows the initial magnetization curves at various temperatures after correcting for demagnetization effect [24]. The curves exhibit typical features of a type-I superconductor. In the $H$–$T$ phase diagram [Fig. 1(f)], the critical field $H_c$ values (red dots) were obtained from the initial magnetization curves, while $H_p$ values (blue dots) were extracted from the deviation points of the Meissner line. The $H_{p\text{-MFM}}$ values (purple dots) extracted from the MFM imaging confirm the accuracy of the $H_p$ determination method from the $M$-$H$ loops. Both $H_c$ and $H_p$ follow the empirical relation $H(T) = H(0)[1 - (T/T_c)^2]$, which fits the experimental data well, yielding $H_c(0) = 872$ Oe.

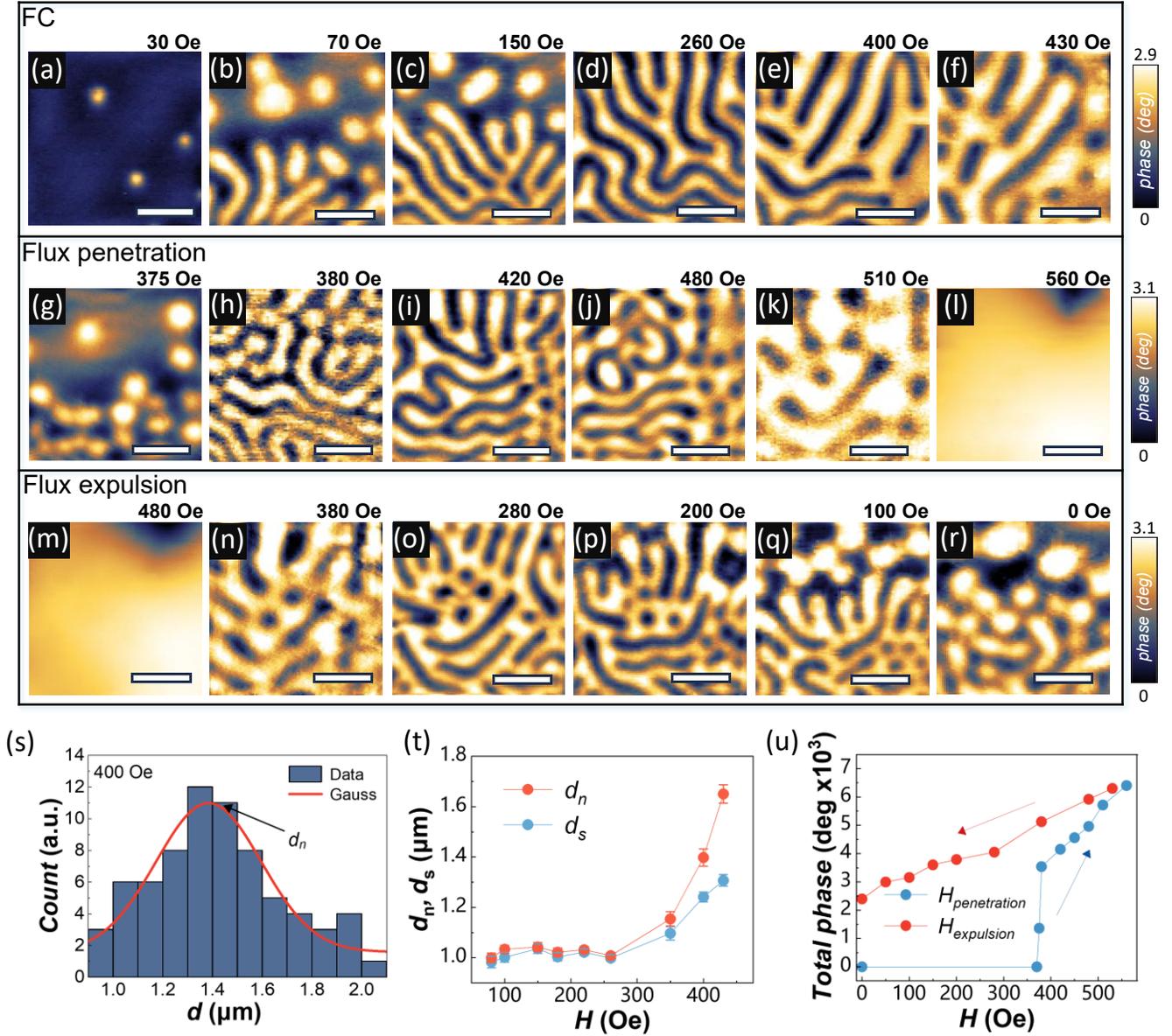

Figure 2. (a)-(f) MFM images of the Ta single crystal obtained after FC to 1.6 K under magnetic fields of (a) 30 Oe, (b) 70 Oe, (c) 150 Oe, (d) 260 Oe, (e) 400 Oe, and (f) 430 Oe. The bright and dark regions correspond to the normal and superconducting states, respectively. Scale bar: 5 μm. (g)-(l) MFM images obtained after ZFC to 1.6 K, with the applied magnetic field subsequently increased to (g) 375 Oe, (h) 380 Oe, (i) 420 Oe, (j) 480 Oe, (k) 510 Oe, and (l) 560 Oe. (m)-(r) MFM images obtained after FC in a magnetic field close to the critical field $H_c$, followed by gradually decreasing the applied field to (m) 480 Oe, (n) 380 Oe, (o) 280 Oe, (p) 200 Oe, (q) 100 Oe, and (r) 0 Oe. (s) Statistical histogram of stripe width at 260 Oe, with the red line representing the Gaussian fitting. (t) Normal-state stripes and spacing variation with the magnetic field. $d_n$ and $d_s$ represent the width of the normal stripes and their spacing. Error bars represent standard error. (u) Variation of the total phase signal with the applied magnetic field during the flux penetration and expulsion processes.

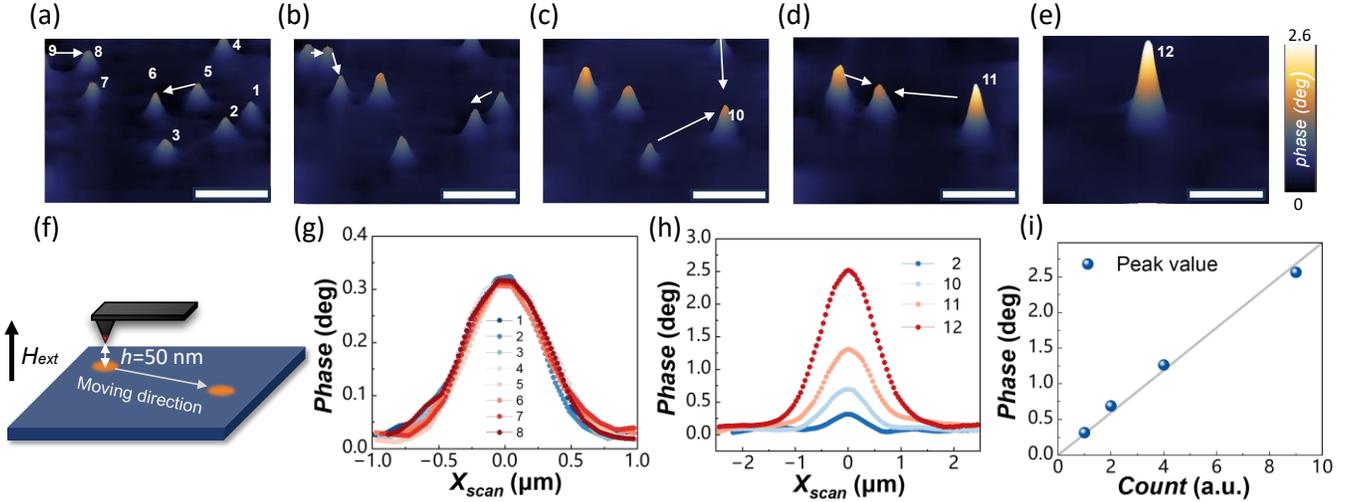

Figure 3. (a) MFM image obtained after field cooling at 1 Oe and 1.6 K. (a–e) Flux tube manipulation demonstrating controlled flux tube merging. Arrows indicate the dragging direction of individual flux tubes. Scale bar: 5 $\mu$m. (f) Schematic illustration of the MFM-tip manipulation process, where flux tubes can be dragged at a tip–sample distance of $h = 50$ nm. (g) Cross-sections of flux tubes after field cooling, suggesting that all of them carry the same amount of flux quantum. (h) Cross-section of a flux tube after merging with different numbers of flux tubes. (i) Peak value of the cross-sectional profiles (dots) and their linear fit (gray line).

Figures 2(a)–2(f) show the flux patterns observed at 1.6 K after field cooling under various magnetic fields. At low fields, the normal domains appear as isolated flux tubes, and the nonuniformity in their diameters indicates that they carry different numbers of flux quanta. With increasing magnetic field, both the diameter and the density of the flux tubes increase. Once the circular flux domains exceed a critical size, the competition between the short-range attraction arising from the positive superconducting–normal interfacial energy and the long-range magnetic interaction drives a spontaneous transformation into stripe-like or dendritic structures [25]. At 70 Oe [Fig. 2(b)], the coexistence of flux tubes and stripe domains is clearly visible, and at 150 Oe [Fig. 2(c)] the stripes extend further and evolve into dendritic morphologies. The stripe domains exhibit a pronounced evolution with increasing magnetic field. To quantitatively analyze this evolution, cross-sectional profiles were taken every 1 $\mu$m along the stripe domains, and histograms of stripe width and spacing were fitted

with Gaussian functions [Fig. 2(s)]. The peak values of the fitted distributions were taken as the characteristic width and spacing at each magnetic field. The resulting magnetic field dependence of stripe width and spacing is shown in Fig. 2(t). Both parameters remain constant around 1 $\mu$m. Comparing the images at 70 Oe and 150 Oe confirmed that the stripe domains were still in the stage of elongation and growth. As the magnetic field increased to 350 Oe, the stripes extended across the entire sample and began to merge, leading to a simultaneous increase in both stripe width and spacing. When the magnetic field is further raised to 430 Oe [Fig. 2(f)], the stripes continue to broaden while their spacing increases slowly; meanwhile, the superconducting stripe domains begin to shrink, gradually transforming into isolated superconducting bubbles. Near the critical field $H_c$, the normal regions eventually occupy the entire sample.

During flux penetration in type-I superconductors, the flux is hindered by the geometric barrier [10, 11], which leads to the formation of finger-like normal extensions at the sample edges [12]. Flux tubes can detach from these extensions and rapidly move toward the sample center driven by Lorentz forces from the Meissner current [13]. As shown in Fig. 2(g), when the applied magnetic field increases to 375 Oe, the flux tubes appear. This value is slightly higher than the $H_p = 367$ Oe measured at 1.8 K obtained from Fig. 1(f), which is reasonable since the penetration field $H_p$ measured at 1.6 K should be slightly larger. This agreement further validates the reliability of our $H_p$ determination method. The flux tubes observed at 375 Oe [Fig. 2(g)] exhibit nonuniform sizes, which can be attributed to local variations in the energy barrier caused by minor edge inhomogeneities [26], leading the flux tubes to carry different numbers of flux quanta [27]. As the magnetic field increases, the flux tubes become densely packed; when their spacing becomes too small, the positive interfacial energy between the superconducting and normal regions dominates, resulting in the merging of flux tubes into stripes at 380 Oe [Fig. 2(h)]. Upon further increasing the field to 420 Oe [Fig. 2(i)], the flux stripes coalesce, giving rise to bubble-like superconducting domains, similar to those observed at 430 Oe in Fig. 2(f).

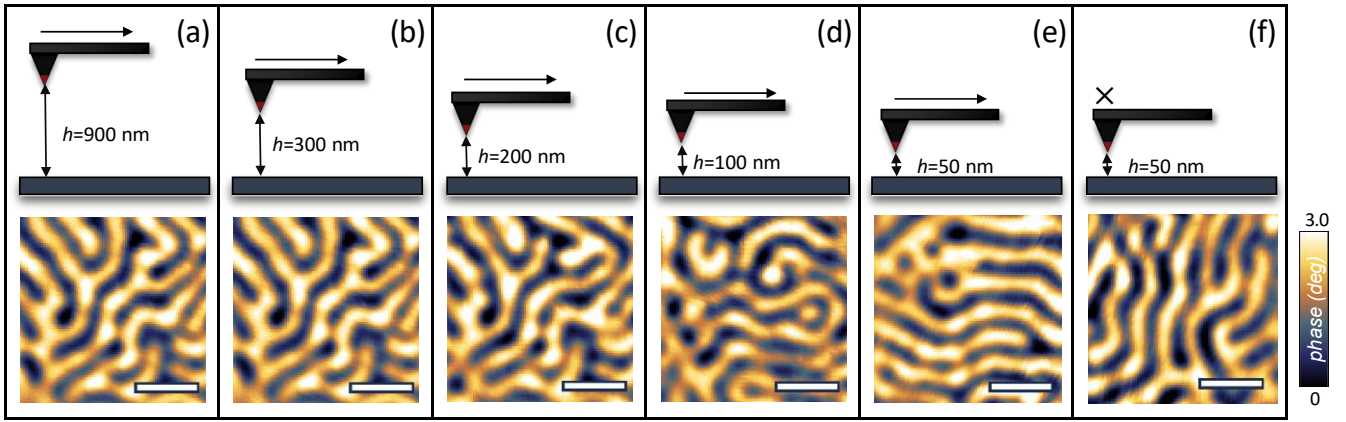

Figure 4. (a) MFM image obtained after field cooling at 220 Oe. (b-e) Images showing controlled manipulation of flux stripes. After being laterally dragged by the MFM tip, the magnetic flux stripes are arranged along the scanned direction. (f) Resulting longitudinal arrangement of flux stripes after dragging. Schematic illustration above each image showing the tip-sample height and moving direction. Scale bar: 5 $\mu$m.

Figures 2(m)–2(r) illustrate the flux expulsion process upon gradually decreasing the magnetic field from near $H_c$. In contrast to flux penetration, the flux stripes become progressively narrower as the field decreases, and some stripes further decompose into flux tubes. Even at zero field [Fig. 2(r)], residual flux tubes and shortened stripes remained, which could be ascribed to pinning effects [28]. Since the sample surface is flat, these pinning centers likely originate from bulk pinning formed during crystal growth. Moreover, at 100 Oe during flux expulsion [Fig. 2(q)], superconducting bubbles are still observed, suggesting that the formation of such structures is not solely driven by the compression of superconducting domains at high fields; rather, similar bubble states can also emerge from stripe motion at low fields.

As shown in Fig. 2(u), we integrated the phase signals of the normal-state regions during the flux penetration (blue dots) and expulsion (red dots) processes. The results show that within the same magnetic field range, the integrated phase value of the flux expulsion process is significantly higher than that of the penetration process, indicating a substantial difference in the total magnetic flux. This provides strong evidence at the microscopic scale for the existence of magnetic hysteresis in the Ta single crystal. The corresponding images clearly reveal the differences in the flux structures: at 380 Oe [Fig. 2(h)], the penetration process exhibits a transition from flux tubes to stripes, whereas during expulsion at the same field [Fig. 2(n)], broad stripe domains persist. Similarly, at 480 Oe [Fig. 2(j)], the penetration process already shows a striped configuration, while the expulsion process [Fig. 2(m)] remains dominated by extended normal regions. The large discrepancies in both the topological structures and the total phase signals between the two processes indicate the presence of topological hysteresis [29, 30]. This topological hysteresis, together with the pinning-induced flux retention during field reduction, accounts for the magnetic hysteresis observed in the magnetization loops in Fig. 1(d).

In type-I superconductors, the interplay between long-range repulsion and short-range attraction leads to the spontaneous formation of multi-quantum flux tubes (giant vortices) [Fig. 2(g)]. However, the fusion process of such flux tubes has not yet been directly observed. Inspired by vortex manipulation techniques in type-II superconductors [17, 18], we attempted to achieve controlled fusion of flux tubes via local MFM manipulation. Under a field-cooled condition of 1 Oe [Fig. 3(a)], isolated flux tubes were imaged when the MFM tip was positioned 900 nm above the sample surface. Reducing the tip–sample distance to 50 nm enhanced the attractive interaction between the tip and the flux tubes, enabling precise manipulation. As shown in Fig. 3(f), the tip was aligned over a selected flux tube and dragged along the indicated direction, moving flux tube-5 toward flux tube-6 and flux tube-9 toward flux tube-8. The merged flux tubes exhibited a markedly larger diameter [Fig. 3(b)], while flux tube-9 became trapped by a strong pinning center during motion. Repeating this process [Figs. 3(c)–3(d)] resulted in the complete fusion of all nine flux tubes [Fig. 3(e)]. Cross-sectional phase profiles of the individual flux tubes [Fig. 3(g)] reveal nearly identical peak heights of 0.3 deg, indicating that each tube carries the same amount of flux quanta after field cooling. Figure 3(h) shows the cross-sectional analysis of merged flux tubes with different numbers, and the peak values extracted from these profiles are presented in Fig. 3(i). The peak value increased linearly with the number of merged flux tubes, with a slope of approximately 0.3 deg per count, confirming that the flux tubes in Fig. 3(e) are indeed formed by intentional manipulation and merging.

In addition to manipulating flux tubes, we also achieved controlled reconfiguration of flux stripe patterns using the MFM tip. Initially, the sample was field-cooled under an applied field of 220 Oe to obtain the original stripe configuration [Fig. 4(a)]. The tip–sample distance was then gradually reduced, and lateral dragging experiments were performed at reduced heights [Figs. 4(b)–4(e)]. The upper panels of Figs. 4(a)–4(f) illustrate the MFM tip manipulation process: the tip was first moved forward at indicated height, and then lift off 900 nm before moving backward. After each line of dragging, the tip was displaced 500 nm, and the process was repeated 30 times, covering the entire scanned region. This process makes sure that the manipulation of the strip is always along the same direction. Finally, the tip was lifted back to 900 nm to image the reconstructed flux stripes. The experimental results revealed a progressive reorientation of the stripe domains as the tip approached the surface. When the tip was lifted 300 nm [Fig. 4(b)], the magnetic attraction was too weak to overcome the pinning effect and the interaction between flux stripes. When the tip height was reduced to 200 nm [Fig. 4(c)], partial deformation of the stripes became visible. At 100 nm [Fig. 4(d)], the flux stripes were largely reconfigured into a transversely aligned pattern, and at 50 nm [Fig. 4(e)] they became fully reoriented along the lateral direction. To verify the reproducibility of the experiment, the sample was field-cooled again under the same magnetic field, and after longitudinal dragging, the magnetic flux lines could be adjusted to a longitudinal arrangement [Fig. 4(f)].

The flux structures in the IS could be easily frozen in a metastable state due to intricate interactions. The stability of flux stripes is studied to reveal any structure transformations through an AC force perturbation. To address this, the sample was first field-cooled at 220 Oe. Then, an AC current with frequency of 10 Hz was applied with gradually increasing its amplitude from 16 mA to 25 mA [Figs. 5(a)-5(f)]. We found that at an AC current of 21 mA, the flux stripes reorganized into a grid-like pattern [Fig. 5(c)], where the originally continuous superconducting stripes fragmented into discrete superconducting bubbles arranged in a lattice reminiscent of the vortex lattice in type-II superconductors. Such bubble lattices typically appear at high magnetic fields [16, 31], and similar local patterns were also observed in Fig. 2(j) during flux expulsion at high fields. Further increasing the AC current to 25 mA [Fig. 5(f)] caused the flux pattern to revert to a stripe configuration, indicating that this transition was not merely due to simple flux motion. Moreover, the emergence of the grid-like pattern was accompanied by a pronounced increase in the area of the normal regions, suggesting an overall increase in magnetic flux under AC driving. To confirm this, we integrated the phase signal over the flux regions, which revealed a significant increase in the total value after the stripe-to-grid transition [Fig. 5(g)], confirming the enhanced flux density.

The observed increase of the normal-phase fraction can only arise from flux penetration from the sample edges [32]. Microscopic inhomogeneities at the sample edge lead to spatial variations in the local geometric barrier [27]. Consequently, under the combined action of AC dynamic perturbations and the Lorentz force, flux tubes preferentially penetrate the sample at positions where the local barrier is relatively lower [33]. Once inside, the flux tubes moved further into the sample interior under the combined action of the Lorentz force and interactions with neighboring flux domains, eventually slowing down and accumulating at pinning centers [34]. Flux moving toward the sample edges was impeded by the geometric barrier [35], preventing its exit from the sample and leading to an increase in flux density. This is consistent with the observations in Fig. 2(r), where residual flux remained even at zero field, demonstrating that pinning potentials and geometric barriers jointly impede flux expulsion. At this high flux density, energy minimization drove the system to reorganize into superconducting bubble structures [16], corresponding to the flux-grid state observed experimentally. When the AC current was further increased to 25 mA [Fig. 5(f)], the Lorentz force overcomes both pinning and geometric barriers, allowing excess flux to exit and the pattern reverting to the stripe state.

To gain deeper insight into the dynamics of flux stripes under AC excitation, we systematically studied the evolution of flux patterns as a function of AC current amplitude under different magnetic fields and frequencies. The resulting phase diagram of flux structures [Fig. 5(h)] revealed that the threshold current required to enter or exit the flux-grid state decreases monotonically with increasing magnetic field. This is because, at higher applied magnetic fields, the system approaches the penetration field and the geometric barrier becomes weakened. Consequently, only a relatively small Lorentz force together with AC perturbations is sufficient to drive the magnetic flux to overcome the residual barrier, thereby promoting its penetration and migration within the superconductor. At a fixed magnetic field of 260 Oe, the dependence of flux-pattern evolution on AC frequency [Fig. 5(i)] showed the opposite trend: the threshold current increased monotonically with frequency. As the AC frequency increased, the magnetic flux did not have sufficient time to overcome pinning potentials and geometric barriers before the current reversed. Consequently, a higher current was required to overcome these barriers and drive flux motion.

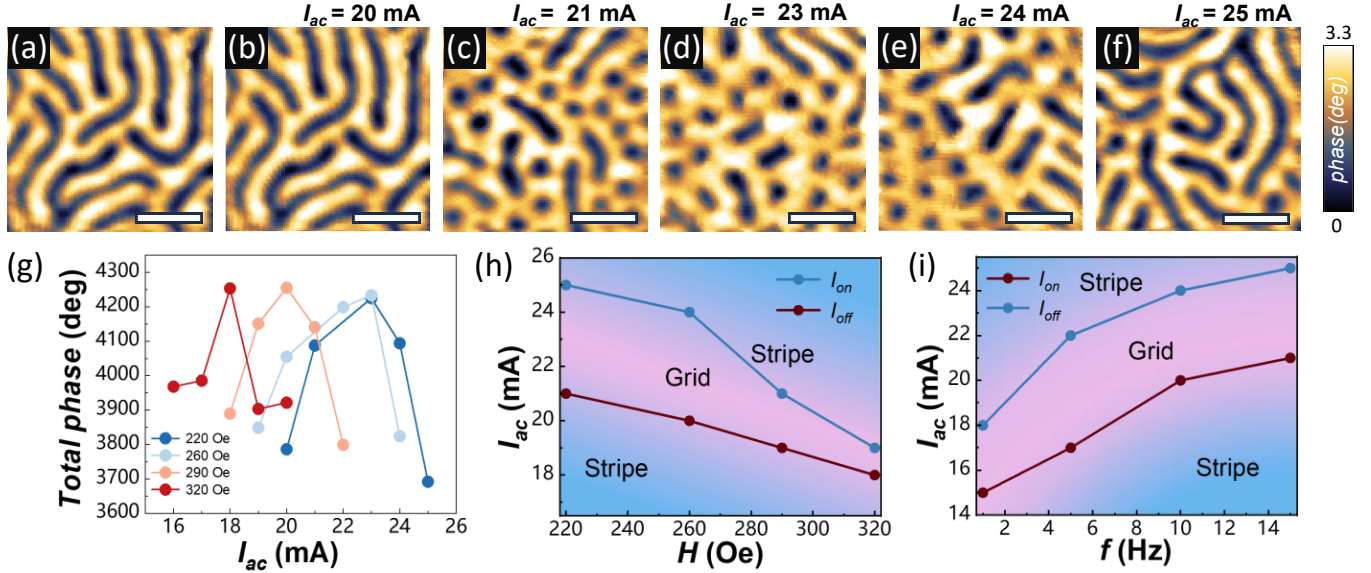

Figure 5. (a) MFM image obtained after field cooling at 220 Oe. Scale bar: 5 $\mu$m. (b)-(f) Flux patterns obtained after applying a 10 Hz AC current with varying amplitudes for 5 min. (g) Variation of the total phase signal of flux regions under different cooling magnetic fields as a function of the applied alternating current amplitude. (h) $I_{ac}$-$H$ phase diagram showing the flux topology transition at 10 Hz. (i) $I_{ac}$-$f$ normal phase diagram showing the flux topology transition at 220 Oe. Here, $I_{on}$ ($I_{off}$) represents the threshold AC current amplitude at which the flux-grid state forms (disappears).

## IV. CONCLUSIONS

To conclude, this study addresses the critical gap in the active control and dynamic understanding of flux domains in the intermediate state of type-I superconductors. By uniquely leveraging the dual capability of low-temperature magnetic force microscopy (MFM) for both high-resolution imaging and local manipulation, we establish a pathway for the controllable reconstruction of flux structures. The controlled fusion of individual flux tubes and reconfiguration of stripe patterns via the MFM tip provide microscopic evidence for interface-energy-driven attraction. The discovery of a reversible, current-driven "stripe-grid-stripe" transition under AC excitation, along with the constructed phase diagram, demonstrates how AC perturbations cooperate with the Lorentz force to modulate flux entry, trapping, and structural reorganization. Ultimately, this work reveals the rich non-equilibrium dynamics of type-I flux structures and demonstrates their active manipulation, paving the way for potential flux-based superconducting devices.

## V. ACKNOWLEDGMENTS

This paper is supported by the National Natural Science Foundation of China (Grants No. 12174242). J.-Y.G. is also grateful for the support by the Program for Professor of Special Appointment (Eastern Scholar) at Shanghai Institutions of Higher Learning.